\pdfoutput=1
\documentclass[aps,prd,amsmath,floats,floatfix, twocolumn,
superscriptaddress,nofootinbib,showpacs,longbibliography]{revtex4-1}
\usepackage[T1]{fontenc}
\usepackage[utf8]{inputenc}
\usepackage{lmodern}
\usepackage{mathrsfs}
\usepackage{amsthm}
\usepackage{verbatim}
\usepackage{makecell}

\usepackage[dvipsnames, usenames]{xcolor}
\definecolor{linkcolor}{rgb}{0.0,0.3,0.5}
\usepackage[hypertexnames=false, unicode, colorlinks=true, linkcolor=linkcolor,
citecolor=linkcolor, filecolor=linkcolor,urlcolor=linkcolor,
pdfusetitle]{hyperref}

\usepackage[all]{hypcap}
\usepackage{graphicx}
\usepackage{xspace}
\usepackage{braket}
\usepackage{amssymb}
\usepackage[normalem]{ulem} %for \sout
\usepackage{bm} % boldmath

\usepackage{array}
\usepackage{multirow}
\usepackage{microtype}

\usepackage{lipsum}  

\usepackage{blindtext}
\usepackage{mathrsfs}
\usepackage{orcidlink}
\graphicspath{%
  {figs/}%
}

\DeclareMathAlphabet{\mathpzc}{OT1}{pzc}{m}{it}

\newcommand{\red}{}
\usepackage{mathtools}
\def\scri{\mathscr{I^+}}
\def\H{\mathscr{H^+}}

\newcommand{\half}{\frac{1}{2}}

\newtheorem{thm}{Theorem}[section]
\newtheorem{lmm}[thm]{Lemma}

\newtheorem{defi}{Definition}[section]

\newtheorem{rem}{Remark}[section]

\newcommand{\UMassDMath}{\affiliation{Department of Mathematics, University of Massachusetts, Dartmouth, MA 02747, USA}}
\newcommand{\UMassDPhysics}{\affiliation{Department of Physics, University of Massachusetts, Dartmouth, MA 02747, USA}}
\newcommand{\CSCDRUMass}{\affiliation{Center for Scientific Computing and Data Science Research, University of Massachusetts, Dartmouth, MA 02747, USA}}
\newcommand{\KTH}{\affiliation{Department of Mathematics, KTH Royal Institute of Technology, Stockholm, Sweden.}}

\newcommand{\TITLE}{Superconvergent Discontinuous Galerkin Method for the Scalar Teukolsky Equation on Hyperboloidal Domains: Efficient Waveform and Self-Force Computation}
\begin{document}

\title{\TITLE}

% \author{Manas Vishal} \email{vishalmanas28@gmail.com} \UMassDMath \CSCDRUMass
% \author{Scott E. Field\,\orcidlink{0000-0002-6037-3277}} \UMassDMath \CSCDRUMass
% \author{Sigal Gottlieb\,\orcidlink{0000-0002-6526-3886}} \UMassDMath \CSCDRUMass
% \author{Jeniffer Ryan\,\orcidlink{0000-0002-6252-8199}} \KTH
% \author{Gaurav Khanna} \URI \UMassDPhysics \CSCDRUMass
%% For some reason Gaurav's orcid causes latex to throw errors... but it will still compile the PDF
% \author{Gaurav Khanna \orcidlink{0000-0002-2565-8170}} \URI \UMassDPhysics \CSCDRUMass

%%%%%%%%%%without ORCID
\author{Manas Vishal\,\orcidlink{0000-0003-3424-3505}} \email{vishalmanas28@gmail.com} \UMassDMath \UMassDPhysics \CSCDRUMass
\author{Scott E. Field\,\orcidlink{0000-0002-6037-3277}} \UMassDMath \CSCDRUMass
\author{Sigal Gottlieb\, \orcidlink{0000-0002-6526-3886}} \UMassDMath \CSCDRUMass
\author{Jennifer Ryan \orcidlink{0000-0002-6252-8199}}  \KTH
\date{\today}

%==========================================================================
\begin{abstract}
  The long-time evolution of extreme mass-ratio inspiral systems requires minimal phase and dispersion errors to accurately compute far-field waveforms, while high accuracy is essential near the smaller black hole (modeled as a Dirac delta distribution) for self-force computations. Spectrally accurate methods, such as nodal discontinuous Galerkin (DG) methods, are well suited for these tasks. Their numerical errors typically decrease as \(\propto (\Delta x)^{N+1}\), where \(\Delta x\) is the subdomain size and \(N\) is the polynomial degree of the approximation. However, certain DG schemes exhibit superconvergence, where truncation, phase, and dispersion errors can decrease as fast as \(\propto (\Delta x)^{2N+1}\). Superconvergent numerical solvers are, by construction, extremely efficient and accurate. We theoretically demonstrate that our DG scheme for the scalar Teukolsky equation with a distributional source is superconvergent, and this property is retained when combined with the hyperboloidal layer compactification technique. This ensures that waveforms, total energy and angular-momentum fluxes, and self-force computations benefit from superconvergence. We empirically verify this behavior across a family of hyperboloidal layer compactifications with varying degrees of smoothness. Additionally, we show that \red{dissipative} self-force quantities for circular orbits, computed at the point particle's location, also exhibit a certain degree of superconvergence. Our results underscore the potential benefits of numerical superconvergence for efficient and accurate gravitational waveform simulations based on DG methods. 
\end{abstract}

\maketitle

%==========================================================================

%%%%%%%%%%%%%%%%%%%%%%%%%%%%%%%%%%%%%%%%%%%%%%%%%%%%%%%%%%%%%%%%%%%%%%%%%%%%%%%
\section{Introduction}
\label{Sec:Introduction}

High-order numerical methods have become indispensable for solving partial differential equations (PDEs) that arise in general relativity and gravitational wave modeling. Many of these problems share a similar set of features and goals. For example, computing the far-field waveform is essential, as it is directly observed by gravitational wave detectors. Near-field quantities can also be important; for instance, in black hole perturbation theory applied to extreme mass-ratio inspirals (EMRIs)~\footnote{An EMRI is comprised of a \red{compact object of a small mass--$q$} orbiting a large mass--$M$ \red{compact object}, where $q \ll M$. EMRI systems emit gravitational radiation at low frequencies and are a key target of the upcoming Laser Interferometer Space Antenna (LISA) observatory~\cite{AmaroSeoane:2017las,Amaro-Seoane_2007,2010arXiv1009.1402A}.}, a key measurement is the self-force, computed in the vicinity of the smaller black hole. \red{EMRI waveform modeling relies crucially on black hole perturbation theory and self-force computations.~\cite{afshordi2023waveform,abac2025science}}.

Discontinuous Galerkin (DG) methods~\cite{reed1973triangular,Hesthaven2008,cockburn2000development,cockburn2001devising,cockburn1998runge} are a class of high-order numerical schemes that offer spectral accuracy and other favorable, problem-specific attributes. For example, a standard approach for studying EMRIs is to approximate the smaller compact object as a point--like Dirac delta distribution. DG methods \red{(posed as an integral form of the governing equations)} can incorporate distributional source terms proportional and their derivatives \red{without any functional approximation of the source term on the numerical grid}~\cite{fan2008generalized,field2009discontinuous,field2022discontinuous}, allowing the smaller black hole to be accurately modeled. As shown in previous work, DG methods for such problems can be designed such that numerical errors decrease as \(\propto (\Delta x)^{N+1}\), where \(\Delta x\) is the subdomain size and \(N\) is the polynomial degree of the approximation. This convergence rate holds throughout the numerical grid, including at the location of the Dirac delta distribution, where the solution may exhibit discontinuities.

All numerical methods, including DG methods, face challenges on unbounded domains where waves propagate to infinity. Hyperboloidal layers~\cite{zenginouglu2011hyperboloidal,zenginouglu2011null} provide an elegant solution to the challenge of imposing correct boundary conditions and extracting the far-field waveform. By smoothly transitioning the computational domain to future null infinity, these layers enable the direct extraction of gravitational wave signals as they would appear at infinity, while eliminating the need for artificial boundary conditions that can introduce spurious reflections. 

Numerical schemes combining spectrally convergent DG methods with hyperboloidal layers are especially attractive due to their efficiency and accuracy. Interestingly, as we describe in this paper, the combination of these two techniques can lead to a superconvergent method, that is, a method that converges faster than the theoretically predicted rate of \(\propto (\Delta x)^{N+1}\) expected from polynomial interpolation error estimates. For DG schemes that exhibit superconvergence, the truncation, phase, and dispersion errors can decrease~\footnote{The $2N+1$ superconvergent rate is for hyperbolic equations, for elliptic and parabolic problems it is one order lower~\red{\cite{adjerid2002posteriori,baccouch2014superconvergent}}.} as fast as \(\propto (\Delta x)^{2N+1}\). This provides even greater computational benefits. \red{We suspect that this observation will generalize to alternative, scri-fixing hyperboloidal foliations \cite{Zenginoglu:2007jw,Zenginoglu:2009ey,Ansorg:2016ztf,Jaramillo:2020tuu}, although we do not consider that possibility here.}

Superconvergent methods, originally discussed by Biswas et al. \cite{biswas1994parallel}, have long been an area of interest in numerical analysis. The phenomenon has since been rigorously studied in various contexts, including elliptic, parabolic, and hyperbolic equations \cite{adjerid1995high, adjerid2007discontinuous, adjerid2009discontinuous, baccouch2014superconvergent}. Superconvergence in DG methods can manifest at specific points within a subdomain, such as Radau nodal points, where the error can decrease at an accelerated rate \cite{adjerid2012superconvergent, cheng2008superconvergence, ji2012accuracy} depending on the choice of numerical flux~\cite{qiu2005discontinuous,meng2016optimal}. In addition, post-processing techniques, such as the Smoothness-Increasing Accuracy-Conserving (SIAC) filter \cite{ji2012accuracy}, can further extract superconvergence at arbitrary grid points.

In a previous study~\cite{Vishal:2023fye}, and motivated by EMRI modeling, we developed a DG for solving the distributionally-sourced, $s=0$ Teukolsky equation describing scalar waves on Kerr. By extending the numerical flux construction of Refs.~\cite{fan2008generalized,field2009discontinuous,field2022discontinuous} to this new system and applying the hyperboloidal layer technique, we empirically observed that our DG scheme exhibited superconvergence. This unexpectedly pleasant finding suggested that DG methods -- especially combined with hyperboloidal layers -- may be more efficient for gravitational wave simulations than previously thought.  Building on these initial observations, this paper provides a more detailed discussion of superconvergent DG methods on hyperboloidal domains. \red{We note that related DG methods have been explored in other previous works, and the supercovergent techniques we discuss here could also be applied to those methods \cite{diener2012self,cupp2022equal}.}

In this paper, we establish when and how our DG scheme exhibits superconvergence, explaining why this property emerges and under what conditions it can be expected. We also perform numerical experiments to confirm that superconvergence holds in practical simulations relevant to EMRI modeling. Notably, we show that waveforms extracted at future null infinity, at the horizon, \red{and dissipative component of self-force quantity} computed at the small body's location benefits from superconvergence, enabling highly accurate results on comparatively coarse computational grids. To our knowledge, this work represents the first explicit demonstration of superconvergence within the context of numerical relativity. Our findings suggest that DG methods, and more broadly superconvergent numerical schemes, could enhance the efficiency and accuracy of gravitational wave modeling especially when combined with the hyperboloidal layer technique.

The rest of this paper is structured as follows. Section~\ref{sec:preliminaries} provides background on hyperbolic PDEs, hyperboloidal layers, and DG methods. Section~\ref{sub:superconvergence_in_DG} examines superconvergence in DG methods and its implications for numerically computed gravitational waveforms on hyperboloidal domains. Section~\ref{sec:numerical_experiments} presents numerical examples of superconvergence, including waveforms from the ordinary wave equation, the scalar Teukolsky equation, and self-force calculations for circular orbits.

\section{Preliminaries}
\label{sec:preliminaries}

\subsection{Hyperbolic PDEs using hyperboloidal layers}
\label{Sec:Hyperbopolic_eqs}

In this paper, we will consider coupled systems of linear, hyperbolic equations in one-dimension that can be written in the form 
\begin{align} \label{eq:matrix_vector_system}
\dot{U}+\hat{A}U^{\prime}+\hat{B}U = \hat{G}(t) \delta(x - x_p ) \,,
\end{align}
where the prime stands for differentiation with respect to the spatial variable $x \in [a,b]$, 
the dot stands for differentiation with respect to the temporal variable $t$, 
$U(t,x)$ is a solution (column) vector of length $L$,
and $\hat{B}(x)$ is an $L$-by-$L$ matrix.
We consider differential equations sourced by a Dirac delta distribution
$\delta(x - x_p)$ located at $x=x_p$, where $\hat{G}(t)$ is a vector of length $L$ whose components are prescribed functions of time.
The matrix $\hat{A}(x)$ is of size $L$-by-$L$ and, for the hyperbolic systems we will consider,
it is diagonalizable with real eigenvalues.

The numerical simulation of wave phenomena on an open domain requires the specification of radiation boundary conditions and, in many cases of practical interest, access to the far-field waves are of particular importance. 
However, unless the matrices $\hat{A}$ and $\hat{B}$ take a particularly simple form, specification of the correct boundary conditions and extraction of the far-field wave signal can be challenging. One approach is to derive non-reflecting boundary conditions and near-field-to-far-field kernels~\cite{hagstrom1999radiation,hagstrom2007radiation,lau2005analytic,lau2004rapid,grote1996nonreflecting,Benedict:2012kw,Field:2014cka}. Another approach, and the one we pursue here, is to use the method of hyperboloidal layers~\cite{zenginouglu2011hyperboloidal,zenginouglu2011null,Harms:2013ib}.

We first introduce hyperboloidal coordinates, $(\rho, \tau)$, defined by
\begin{align} \label{eq:hyperboloidal_transformation}
x=\frac{\rho}{\Omega(\rho)} \,, \qquad \tau=t-h(x) \,,
\end{align}
that are related to the original $(x,t)$ coordinates through specification of the functions,
\begin{align} \label{eq:compression_and_height}
 & \Omega=1-\left(\frac{\rho-R}{s-R}\right)^{P}\Theta(\rho-R) \,, \quad  h=\frac{\rho}{\Omega}-\rho \,,
\end{align}
where $R$, $s$, and $P$ are parameters of the coordinate transformation
and $\Theta$ is the Heaviside step function. 
To the left of $\rho=R$, the coordinates are
the original $(x,t)$ ones. 
To the right of $\rho=R$, the coordinates smoothly connect the computational domain to 
future null infinity, which is defined by $\rho=s$ (formally $x=\infty$).
We will choose the location of the interface, $R$, such that 
(i) the source term is always located to the left of the interface and (ii) for our multi-domain method, we collocate
the start of the layer at a subdomain interface. A sufficiently smooth coordinate transformation can be achieved by
setting the value of $P$, typically taken to be a positive integer~\cite{bernuzzi2011binary}. The value of $P$ could potentially impact the numerical scheme's order of convergence. \red{We study the dependence of the error on $P$ in Section.~\ref{sub:ordinary_wave_equation} and in our previous work~\cite{Vishal:2023fye}}. For our multi-domain method, however,
we anticipate that so long as $H$, given by Eq.~\eqref{eq:HypLayers}, is continuous at a subdomain's interface our DG method's convergence properties will not be impacted. This expectation was confirmed in Ref.~\cite{Vishal:2023fye}.

Noting that $\partial_{\tau} = \partial_t$ and $\partial_x = (1-H) \partial_{\rho} - H \partial_{\tau}$,
it is straightforward to show that in hyperboloidal coordinates, Eq.~\eqref{eq:matrix_vector_system} becomes
\begin{align} \label{eq:matrix_vector_system_hlayers}
\dot{U}+AU^{\prime}+BU = G(t) \delta(x - x_p ) \,,
\end{align}
where~\footnote{While the hyperboloidal coordinate transformation is singular at $\rho=s$, 
the coefficients of $A$ and $B$ must be well behaved in the limit $\rho \rightarrow s$. 
This is certainly the case for the systems considered in
Sec.~\ref{sec:numerical_experiments}, while general conditions of 
applicability are discussed in Ref.~\cite{zenginouglu2011hyperboloidal}.}
$A = \left(\mathbf{I}-HA\right)^{-1}\hat{A}\left(\mathbf{I}-H\right)$,
$B = \left(\mathbf{I}-HA\right)^{-1}\hat{B}$,
$G = \left(\mathbf{I}-HA\right)^{-1}\hat{G}$, and
\begin{align} \label{eq:HypLayers}
H = \partial h / \partial x = 1 - \partial \rho / \partial x = 1 - \frac{\Omega^{2}}{\Omega - \rho \Omega^{\prime}} \,.
%\qquad
%    &= 1 - \omega  \,, \\
%H^{\prime} = \frac{dH}{d\rho}=-\omega^{\prime}  \,,
\end{align}
In Eq.~\ref{eq:matrix_vector_system_hlayers}, 
we now use an over-dot to denote $\partial / \partial \tau$ differentiation
and a prime for differentiation by $\partial / \partial \rho$.
To the left of the layer, where $\rho < R$, we have $H=0$ and so both Eq.~\eqref{eq:matrix_vector_system}
and Eq.~\eqref{eq:matrix_vector_system_hlayers} are identical. One important fact of this 
coordinate transformation is that 
the outgoing characteristic curves obey $\tau - \rho = t - x$. 

Two important properties are achieved by the hyperboloidal coordinate transformation. First, the boundary $\rho=s$ is an outflow boundary and hence no boundary conditions are needed. That is, for the characteristic variables associated with left-moving waves in our original coordinate system (i.e. incoming waves at the right boundary) now have a zero wave speed at $\rho=s$. Second, the location defined by $\rho=s$ is $x=\infty$ in our original coordinate system. That is, gravitational waveforms measured at future null infinity are 
at the coordinate location $\rho=s$.

In the next section we show that when using a hyperboloidal coordinate domain, future null infinity coincides with a superconvergent point on our numerical grid. Furthermore, self-force measurements at $x=x_p$ also happen to be another
superconvergent point on our numerical grid. This allows our scheme to obtain highly accurate waveform and self-force measurements for comparatively sparse computational grids. We note that superconvergence is not a property of hyperboloidal layers, but rather our DG scheme. Yet the two work well together as $\rho=s$ is both a super-convergent point and exactly where we need to record the far-field waveform after applying the method of hyperboloidal layers. Superconvergence in gravitational waveform computations (but not self-forces) were previously reported in Ref.~\cite{Vishal:2023fye}. Yet very little justification was given in that work for why or under what general conditions this can be expected when using hyperboloidal layers. The Sec.~\ref{sub:superconvergence_in_DG} seeks to clarify this.

\subsection{The discontinuous Galerkin method}

We use a standard discontinuous Galerkin method to discretize the distributionally sourced system~\eqref{eq:matrix_vector_system_hlayers}. Our numerical scheme is based on the one described in Refs.~\cite{field2009discontinuous,field2022discontinuous,Vishal:2023fye} for solving one-dimensional wave-like equations written in fully first-order form with source terms proportional to a Dirac delta distribution and its derivative(s). As such, we only summarize the key steps in the discretization and refer interested readers to those references for the details. 

\begin{figure}[htbp]
  \centering
  \includegraphics[width=0.45\textwidth]{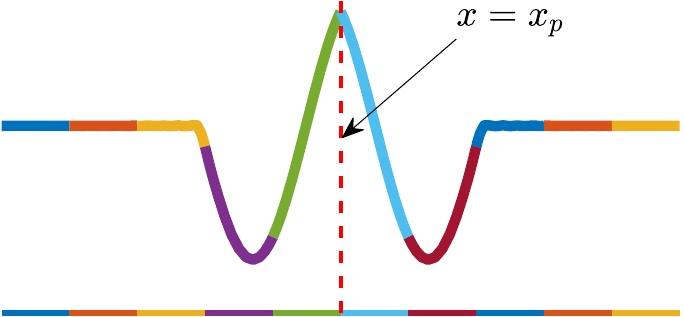 }
   \caption{Cartoon of the solution to the wave equation with a delta source term.
    The Dirac delta distribution source term is located at an interface of two DG subdomains (shown at the bottom of the illustration). Later on in Sec.~\ref{sub:scalar_teukolsky} we show that quantities (such as the self-force) extracted at $x=x_p$ exhibit different convergence and accuracy properties depending on whether the solution on the left or right of $x=x_p$ is used. In particular, the solution evaluated at the left of the Dirac distribution coincides with a Radau outflow nodal point and is found to be orders-of-magnitude more accurate.
    }
  \label{fig:wave_snapshot_dg}
\end{figure}

We partition the spatial domain into $K$ non-overlapping subdomains defined by the partition points $\rho_0 < \rho_1 < \dots < \rho_K = s$ and denote $\mathsf{D}^j = [\rho_{j-1},\rho_j] $ as the $j^{\rm th}$ subdomain. In this one-dimensional setup, the points $\{\rho_i\}_{i=1}^{K-1}$ locate the internal subdomain interfaces, and we require one of them to be the location of the Dirac distribution. Since we are solving a one-dimensional problem, the subdomains are line segments and neighboring subdomains will intersect at a point. A cartoon of our grid is shown in Fig.~\ref{fig:wave_snapshot_dg}.

In each subdomain, all components of $U$, $A$, and $B$ are expanded in a polynomial basis, which are taken
to be degree-$N$ Lagrange interpolating polynomials $\{ \ell_i(\rho) \}_{i=0}^N$ defined from Legendre-Gauss-Lobatto (LGL) nodes. The time-dependent coefficients of this expansion are the unknowns we solve for. Products of terms arising in Eq.~\eqref{eq:matrix_vector_system_hlayers} are represented through pointwise products of the interpolating polynomials at the LGL nodal points.

On each subdomain, we follow the standard DG framework by requiring the residual to satisfy 
\begin{equation} \label{eq:DG_residual}
\int_{\mathsf{D}^j} \left[ \partial_{\tau}  U^j+ A^j \partial_{\rho} U^j + B^j U^j \right] \ell_i^j d\rho  = 
\left[ \left(F^j - F^*\right) \ell_i^j \right]_{x_{j-1}}^{x^j}  \,,
\end{equation}
for all $i=1,2,\dots, N+1$ basis functions. Here we use a superscript ``$j$'' for vectors and matrices whose components have been approximated by Lagrange interpolating polynomials on subdomain $\mathsf{D}^j$. 
Eq.~\eqref{eq:DG_residual} features the physical flux vector, 
%$F(U)=AU$ 
$F^j=A^jU^j$,
and the numerical flux $F^*$.  
The numerical flux is some yet-to-be-specified function $F^*(U^L, U^R)$, where $U^L$ and $U^R$
are, respectively, the left and right boundary values of the numerical solution
defined on the interface between neighboring subdomains. Our numerical flux choice for the distributionally-source system is described in Ref.~\cite{Vishal:2023fye,field2009discontinuous}. 
After carrying out a spatial discretization using the nodal DG method, we integrate the resulting system~\eqref{eq:DG_residual} in time using a fourth-order Runge Kutta (RK4) scheme. The global solution is taken to be a direct sum of the local solutions defined on each subdomain.

\section{Superconvergence in discontinuous Galerkin methods}
\label{sub:superconvergence_in_DG}

In DG simulations certain grid points within an element may exhibit superconvergence. The reason for this phenomenon is 
linked to the behavior of polynomial interpolation~\cite{adjerid2002posteriori}. Before stating the main theoretical 
results that describe this phenomenon~\cite{adjerid2002posteriori}, we remind the reader of the definition of the Legendre polynomials $P_n$ and the related Radau polynomials $R^\pm_{n}$.

\begin{defi}
The Legendre polynomials are defined by the Rodrigues formula
\begin{equation*}\label{eqn:Rodrigues}
P_n(\xi) = \frac{1}{2^nn!}\frac{d^n}{d\xi^n}\left((\xi^2-1)^n\right),\quad -1\leq\xi\leq 1 \,.
\end{equation*}
They are orthogonal polynomials satisfying
\begin{equation*}
\int_{-1}^1P_n(\xi)P_m(\xi)~\textrm{d}\xi = \frac{2}{2n+1}\delta_{nm},
\end{equation*}
where $\delta_{nm}$ is the Kronecker-delta function.  Their recurrence formula, 
\begin{align*}
P_0(\xi) = &1;\quad P_1(\xi) = \xi; \\
P_{n+1}(\xi) = &\frac{2n+1}{n+1}xP_n(\xi)- \frac{n}{n+1}P_{n-1}(\xi) \,,
\end{align*}
make them computationally attractive.  We further note that at the endpoints their values are
\begin{equation*}
P_n(1) = 1 \,,\qquad P_n(-1) = (-1)^n \,.
\end{equation*}
\end{defi}

\begin{defi}
The left and right Radau polynomials for $k>0$ are defined using the Legendre polynomials
\begin{equation*}
 R^-_{k+1}(\xi) = P_{k+1}(\xi) + P_k(\xi) ,\quad\quad R^+_{k+1}(\xi) = P_{k+1}(\xi) - P_k(\xi) \,,
\end{equation*}
respectively. 
\end{defi}

The $(k+1)^{st}$ Radau polynomial has $k+1$ real distinct roots that lie in the interval $[-1,1]$.
We denote these roots  by 
$$\zeta^+_1 < \zeta^+_2 < \dots <\zeta^+_{k+1} = 1$$ 
and
$$-1 = \zeta^-_1 <\zeta^-_2 < \dots < \zeta^-_{k+1} $$ for $R^+_{k+1}(\xi)$ and $R^-_{k+1}(\xi)$, respectively.
Note that the right Radau polynomial, $R^+_{k+1}$ has a root at $\xi = 1$, 
while the left Radau polynomial $R^-_{k+1}$ has a root at $\xi = -1$,  for all $k\geq 0$;
this is a consequence of the endpoint values of the Legendre polynomials.

To define the Radau polynomials on any interval element $I_i = [x_{i-1/2},x_{i+1/2}]$,
we use the mapping $\xi=\frac{2}{h_i}(x-x_{i-1/2})-1$, where $h_i$ is the size of $I_i$. This allows for defining the {\em shifted  Radau polynomials:}
$$R^{\pm}_{k+1,i}(x) = R^{\pm}_{k+1}\left(x_{i-\half}+\frac{h_i}{2}(\xi+1)\right),\quad x \in I_i.$$
We therefore denote the corresponding polynomial roots as
 $$x^{\pm}_{i,j} = x_{i-\half} + \half h_i(\zeta^{\pm}_j+1),\quad j = 1, \dots, k+1.$$

We are finally ready to state the interpolation result which is at the heart of superconvergence. In our case, we only need the right Radau polynomials. \red{In the discussion that follows, we use the lowercase $u$ to denote a general function, as the results are formulated in that context. When compared to the solution vector $U(t,x)$ introduced in Eq.~\eqref{eq:matrix_vector_system}, the function $u$ may be interpreted as representing one component of $U$. It is worth noting that many of the subsequent results are general statements from approximation theory and apply more broadly than to DG methods alone.}

\begin{lmm}\label{lmm:interpolating special}
Let $u \in \mathcal{C}^{k+1}\left(I_i\right),~ i=1,\dots,N$ and $\zeta^{+}_j\in[-1,1],~ j=1,\dots,k+1$, be the roots of 
$R_{k+1}^{+}(\xi)$. Denote the interpolating polynomial on element $I_i$ by
%$$\pi^{+}_iu(x) = \sum_{n=1}^{k+1} L_{n,i}(x),\quad\quad x \in I_i$$
\begin{align}
\pi^{+}_iu(x) = \sum_{n=1}^{k+1} u(x_{i,n}^{+}) L_{n,i}(x),\quad\quad x \in I_i \,,
\end{align}
where
\begin{align*}
L_{n,i}(x) = \prod_{\substack{j=1\\ j\neq n}}^{k+1}\frac{x-x^{+}_{i,j}}{x^{+}_{i,n}-x^{+}_{i,j}} \,,
\end{align*}
is the $k^{\textrm{th}}$ degree Lagrange polynomial interpolating $u$ at the (distinct) roots 
$x_{i,j}^{+} = x_{i-\half}+\frac{h_i}{2}(\zeta_j^{+} + 1), j=1,\dots,k+1,\, i=1,\dots,N$, of the shifted right Radau polynomial $R^{+}_{k+1}(\xi)$ on $I_i$.
Then the interpolation error satisfies
\begin{align}
u(x(\xi)) - \pi^{+}_i u(x(\xi)) =  h_i^{k+1}c_{k+1}R^{+}_{k+1}(\xi) + \sum_{\ell=k+2}^{\infty}Q_{\ell}(\xi)h_i^{\ell},
\end{align}
where $Q_{\ell}(\xi)$ is a polynomial of degree at most $\ell$.
\end{lmm}

This lemma means that while the interpolation error is generally $O(h^{k+1})$, at the $k+1$ special points $x(\zeta^{+}_{i,j})$ the first term in the error drops away (because $R^{+}_{k+1}(\zeta^{+}_j) =0$) and we obtain $O(h^{k+2})$.
In Theorem~\ref{Thm:pointwise}, we will see 
%later 
that even 
%better results
higher convergence rates can be obtained at the right endpoint $\zeta^{+}_{i,k+1} = 1$.

Lemma \ref{lmm:interpolating special} is the key to the superconvergence of the DG method. 
This superconvergence was shown in \cite{CaoYang}, and we restate it here:
 \begin{thm}\label{Thm:pointwise}
 Consider the approximate solution $u_h$ to the one-dimensional linear hyperbolic conservation law
 obtained by a DG scheme  using basis functions less than or equal to degree $k$ on a uniform mesh and the upwind flux. The numerical initial condition on $I_i$ is given by the 
interpolating polynomial $\pi^{+}_i u(x,0)$.

Let $\xi_i = \frac{2}{h_i}(x - x_{i-\half})-1$ be the mapping between the element $I_i$ 
and the canonical element $[-1,1]$. Then the error $e = u - u_h$ satisfies
\begin{eqnarray}
e(\xi,h,t) ~=~\sum_{\ell=k+1}^{\infty}Q_{\ell}(\xi_i,t)h_i^{\ell}, \quad Q_{\ell}(\cdot,t)\in\mathscr{P}^{\ell}([-1,1]),
\end{eqnarray}
where  
\begin{eqnarray*}
Q_{k+1}(\xi_i,t)= c_{k+1}R^{+}_{k+1}(\xi_i) ,
\end{eqnarray*}
for $c_{k+1}$ depends on $u, h, k$ and $t$,
and 
\[Q_{\ell}(1,t)= 0 \; \; \; \mbox{for} \; \; \ell = k+1, k+2, . . . , 2 k +1. \]
\end{thm}

This theorem shows that the error of the DG solution has a similar form to the interpolation error, which is of order $k+2$ 
at the special points $\zeta^+_{i,j}\, j=1,\dots,k$, and of order $2 k+1$ at the outflow point $\xi = \zeta^+_{i,k+1} = 1$.

\begin{rem}\label{rem:end pointwise}
When $u_h(x,0)\bigg|_{I_i} = \pi^{+}_iu_0(x)$ interpolates $u_0(x)$ at the roots of $R^{+}_{k+1}(\xi_i)$ on element $I_i$, the coefficient of the term on the order of $h^{k+1}$ in the series for 
the initial error satisfies
\begin{equation}\label{eqn:inherent condition}
Q_{k+1}(1,0) = 0.
\end{equation}
In the proof of Theorem~\ref{Thm:pointwise}, this relation was extended to $t > 0$.  
For odd $k$, 
\begin{eqnarray*}
Q_{k+1}(\xi,t) ~&=&~ \left[
b_{k+1}P_{k+1}(\xi)  -  b_{k+1}P_k(\xi)\right]
\\
&=&
b_{k+1} R^+_{k+1}(\xi) 
\end{eqnarray*}
\end{rem}

So far, we have described superconvergence as a property arising from the approximation properties of polynomial interpolation. In our DG method, the solution within each subdomain is locally approximated by a polynomial. However, the superconvergence behavior also depends on the choice of the numerical flux used for coupling subdomains. Previous studies \cite{xu2007error,ji2013negative,frean2020superconvergence} have shown that certain numerical flux choices can reduce the optimal superconvergence order of the numerical solution. Specifically, when the approximate solution (a degree $N$ polynomial) is evaluated at a superconvergent outflow point on the numerical grid, the convergence order at this point is \(2N + m\), where \(m \in \{0, 0.5, 1\}\) depends on the flux choice. When accounting for temporal discretization errors and assuming a uniform spatial grid $h_i = \Delta x$, the overall order of convergence at superconvergent points, given the RK4 time stepper used here, is \(\propto \Delta t^4 + (\Delta x)^{2N+m}\). Away from the Dirac delta distribution, we use an upwind numerical flux \cite{Vishal:2023fye,field2009discontinuous}, for which \(m = 1\).

\subsection{Superconvergence of numerically computed gravitational waveforms}
\label{sub:superconvergence_gravitational_waveforms}

Superconvergence and hyperboloidal layers complement each other, particularly in computing gravitational waveforms for detectors located in the far-field. Given the immense distances involved, these detectors are best approximated as being at infinity, which corresponds to \( \rho = s \) in a hyperboloidal domain. When the computational domain is partitioned into subdomains, the point \( \rho = s \) is positioned at the rightmost boundary of the final subdomain, \( \mathsf{D}^K \). This corresponds to a Radau outflow point 
%\( \xi = \zeta^+_{K,k+1} = 1 \), 
where an upwind numerical flux yields the optimal superconvergence rate of \( 2N+1 \). This theoretical prediction is validated by the numerical results in Sec.~\ref{sec:numerical_experiments}.  

Superconvergence for numerically solving wave equations on hyperboloidal domains has practical implications for the efficiency of numerical solvers -- since the waveform at \( \rho = s \) is the primary output of interest, achieving the optimal superconvergence rate at grid point allows for highly accurate waveform extraction with lower resolution numerical grids.

%%%%%%%%%%%%%%%%%%%%%%%%%%%%%%%%%%%%%%%%%%% Numerical Experiments %%%%%%%%%%%%%%%%%%%%%%%%%%%%%%%%%%%%%%%%%%%%%%%%%%
\section{Numerical experiments}
\label{sec:numerical_experiments}

In this section we perform some numerical experiments to demonstrate the benefits of superconvergence. In particular, we consider the scalar wave equation and the scalar Teukolsky equation. In Sec. ~\ref{sub:ordinary_wave_equation}, we show that the solution to the wave equation exhibits superconvergence which is independent of the hyperboloidal compactification parameter $P$. In Sec. \ref{sub:scalar_teukolsky}, we demonstrate that the solution to the scalar Teukolsky equation exhibits spatial superconvergence at the location of the scalar charge, horizon ($\mathcal{H^+}$), and future null infinity ($\scri$).

\subsection{Ordinary Wave Equation}
\label{sub:ordinary_wave_equation}

Our first experiment considers the ordinary wave equation
\begin{equation}
  \left[ -\frac{\partial^2}{\partial t^2} + \frac{\partial^2}{\partial r^2} - \frac{\ell(\ell+1)}{r^2} \right] \psi_{\ell} = 0 \,,
  \label{eq:wave_equation}
\end{equation}
for which an exact solution can be computed. This equation arises from a multipole expansion~\eqref{eq:harmonic_expansion} of the 3-dimensional wave equation in terms of spherical harmonics. It can also be obtained by setting $M=0$ in Eq.~\eqref{eq:teuk1}. Carrying out a first-order reduction in hyperboloidal coordinates, Eq.~\eqref{eq:wave_equation} becomes the system Eq.~(44) of Ref.~\cite{Vishal:2023fye} (which is the one we numerically solve). Our previous study~\cite{Vishal:2023fye} demonstrated superconvergence across different nodal points on the grid (see Figure 1 of Ref.~\cite{Vishal:2023fye}). Here we explore our numerical scheme's superconvergence properties while varying the hyperboloidal compactification parameter $P$ appearing in Eq.~\eqref{eq:compression_and_height}.

We set purely outgoing initial data for $\ell=2$ such that 
\begin{equation}
\label{eq:outgoing_data}
\psi(t,r) = f''(t - r) + \frac{3}{r} f'(t - r) + \frac{3}{r^2} f(t - r) \,,
\end{equation}
where $f(u)$ is given as,
\[
  f(u) = \sin \left[ f_0 (u - u_0) \right] e^{-c (u - u_0)^2} \,,
\]
$u = t - r$ and $f(u)$ defines the purely outgoing multipole solution. For the data used in this paper, we chose $f_0 = 2$, $c = 1$, and $u_0 = -10$. The purely right-moving solution moves toward null infinity, and the solution is recorded at the rightmost point in the grid, corresponding to null infinity using hyperboloidal layers. The relative $L^{2}$-norm error
\begin{align} \label{eq:rel_L2}
E_{N,K}(\rho) = \frac{\int_{0}^T \left| \psi_{\rm num}(\tau,\rho) -  \psi_{\rm exact}(\tau,\rho) \right|^2 d\tau}{\int_{0}^T \left| \psi_{\rm exact}(\tau,\rho) \right|^2 d\tau} \,,
\end{align}
is computed by comparing the numerical solution, $\psi_{\rm num}(\tau,\rho)$, to the exact solution $\psi_{\rm exact}(\tau,\rho)$. For this numerical experiment we set $\rho=s$ in Eq.~\eqref{eq:rel_L2}, corresponding to null infinity in $(\tau,\rho)$ coordinates.
We remind the reader that $K$ is the number of subdomains and $N$ is the degree of polynomial approximation on each subdomain.

\begin{figure}[h]
  \centering
  \includegraphics[width=0.5\textwidth]{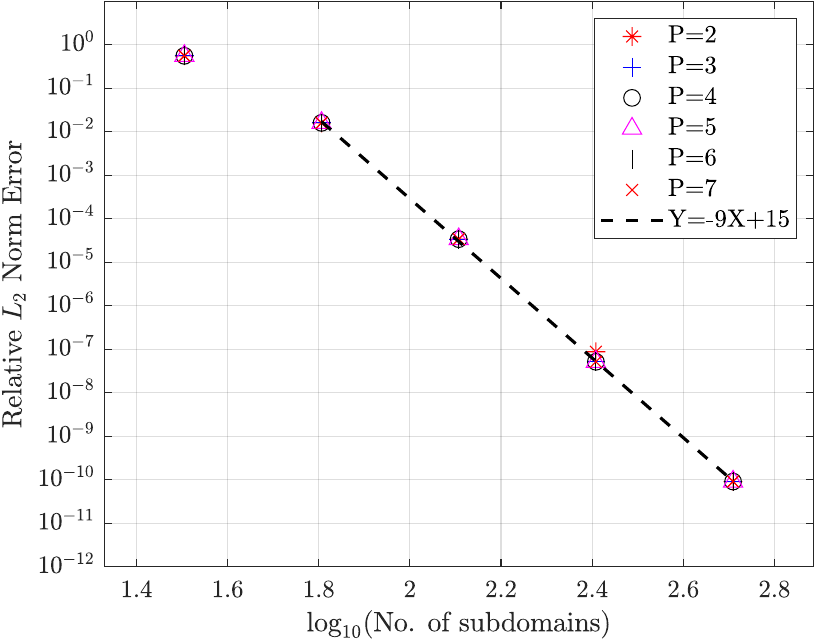}
  \caption{Superconvergence properties of the numerically computed null infinity waveform for the ordinary wave equation~\eqref{eq:wave_equation}. Purely outgoing, $\ell=2$ initial data is set according to Eq.~\eqref{eq:outgoing_data}. The solution is numerically evolved and recorded at the rightmost point in the grid, corresponding to null infinity using hyperboloidal layers. 
  For all cases we set the approximating polynomial's degree to be $N=4$. The relative $L^{2}$-norm is computed according to Eq.~\eqref{eq:rel_L2}, and the error is plotted as a function of the number of subdomains $K$ used in the time-domain solver. We observe that the numerical solution converges to the analytic solution with a rate of $2N+1=9$ for all values of the compactification parameter $P$ of hyperboloidal layer (which controls the smoothness of the coordinate transformation) appearing in Eq.~\eqref{eq:compression_and_height}. While one typically expects a convergence rate of $N+1$ for DG methods (so $N+1=5$ for our setup), our method obtains the theoretically best superconvergence rate of $2N+1$ in the waveform.}
      \label{fig:convergence_with_P}
\end{figure}

Fig.~\ref{fig:convergence_with_P} reports the errors in the solution as a function of $K$ used in the DG solver. As expected 
from the results of Sec.~\ref{sub:superconvergence_in_DG}, the superconvergence properties of the numerical solution does not depend on the value of the hyperboloidal compactification parameter $P$. In particular, we find that the solution converges with a rate of $2N+1$ for all values of $P$, which is faster than the expected rate of $N+1$. For our setup of $\rho \in [1,50]$, we fixed $\ell=2$, $N=4$ and $\Delta t = 10^{-5}$.

\subsection{Scalar Teukolsky Equation}
\label{sub:scalar_teukolsky}

In general relativity, the Teukolsky equation describes the evolution of a scalar, vector, and tensor fields on a Kerr black hole background \cite{teukolsky1972rotating,Teukolsky:1973ApJ}. The master equation governing these fields is given by, 
\begin{eqnarray}
  \label{eq:teuk0}
  &&
  -\left[\frac{(r^2 + a^2)^2 }{\Delta}-a^2\sin^2\theta\right]
           \partial_{tt}\Psi
  -\frac{4 M a r}{\Delta}
           \partial_{t\phi}\Psi \nonumber \\
  &&- 2s\left[r-\frac{M(r^2-a^2)}{\Delta}+ia\cos\theta\right]
           \partial_t\Psi\nonumber\\  
  &&
  +\,\Delta^{-s}\partial_r\left(\Delta^{s+1}\partial_r\Psi\right)
  +\frac{1}{\sin\theta}\partial_\theta
  \left(\sin\theta\partial_\theta\Psi\right)+\nonumber\\
  && \left[\frac{1}{\sin^2\theta}-\frac{a^2}{\Delta}\right] 
  \partial_{\phi\phi}\Psi +\, 2s \left[\frac{a (r-M)}{\Delta} 
  + \frac{i \cos\theta}{\sin^2\theta}\right] \partial_\phi\Psi  \nonumber\\
  &&- \left(s^2 \cot^2\theta - s \right) \Psi = -4 \pi \left(r^2 + a^2 \cos^2\theta \right) T  ,
\end{eqnarray} 
where $M$ and $a$ are the mass and the angular momentum per unit mass of the Kerr black hole, $\Delta = r^2 - 2 M r + a^2$,  
$s$ is the spin weight of the field, and $T$ is the source term driving the perturbation in the background space-time. 

Upon setting $M=1$, $a=0$, and $s=0$, we expand the solution to the scalar (i.e. $s=0$) Teukolsky equation in terms of spherical harmonics to separate the angular part of the solution,
\begin{equation}
\label{eq:harmonic_expansion}
  \Psi(t,r,\theta,\phi) = \frac{1}{r}\sum_{\ell=0}^{\infty} \sum_{m=-\ell}^{\ell} \psi_{\ell m}(t,r) Y_{\ell m}(\theta,\phi) \,.
\end{equation}
By using the properties of the spherical harmonics, we can write the scalar Teukolsky equation as a simple partial differential equation for the coefficients $\psi_{\ell m}(t,r)$,
\begin{equation}
  \label{eq:teuk1}
  \left[ -\frac{\partial^2}{\partial t^2} + \frac{\partial^2}{\partial r_*^2} + V_{\ell}(r) \right]\psi_{\ell m} = S_{\ell m},
\end{equation}
where $r_*=r+2M\ln\left(\frac{r-2M}{2M}\right)$ is the tortoise coordinate,
\begin{eqnarray}
  V_{\ell}(r) = -\left(1-\frac{2M}{r}\right)\left\{ \frac{2M}{r^{3}}+\frac{\ell\left(\ell+1\right)}{r^{2}} \right\}\,,
\end{eqnarray} 
is the effective potential, and $S_{\ell m}$ is a source term we shall specify below.
Carrying out a first-order reduction in hyperboloidal coordinates, Eq.~\eqref{eq:teuk1} becomes the system Eq.~(20) of Ref.~\cite{Vishal:2023fye}, which is the one we numerically solve using trivial initial data $\psi_{\ell m} = \partial_{\tau} \psi_{\ell m} = 0$.

In the standard setup of an extreme-mass-ratio binary black hole problem, the smaller black hole is modelled as a Dirac delta distribution, for which the source term is
\begin{align}
  \label{eq:source_glm2}
  S_{\ell m}(t,r_*) & =
  \frac{-4 \pi q}{u^t \left( r_p^2 + a^2 \right)^{1/2}} \overline{Y}_{\ell m}\left(\frac{\pi}{2},\phi_p(t)\right) \delta(r_{*} - r_{*,p}) \,,
\end{align}
where the constant $r_{*,p}$ denotes the particle's radial position in tortoise coordinates, $\theta_p=\pi/2$ denotes the particle's polar angle, $u^t$ is the t-component of the particle's four-velocity, $q$ is the mass of the smaller black hole,
and $\phi_p(t)$ is the particle's angular location in the $xy$-plane. Expressions for 
\begin{align}
u^t =  g^{t\phi} {\cal L} - g^{tt} {\cal E} \,, \qquad \phi_p(t) = \frac{v^3}{M(1+\tilde{a}v^3)} t\,,
\end{align}
are readily given in, for example, Refs.~\cite{warburton2010self,hughes2000evolution}. Here $v=\sqrt{M/r_p}$, $\tilde{a} = a/M$, and 
\begin{align}
{\cal E} = \frac{1 - 2v^2 + \tilde{a}v^3}{\sqrt{1 - 3v^2 +2\tilde{a}v^3}} \,, \qquad 
{\cal L} = r_p v\frac{1 - 2\tilde{a}v^3 + \tilde{a}^2v^4}{\sqrt{1 - 3v^2 +2\tilde{a}v^3}} \,,
\end{align}
are, the particle's energy and angular momentum, respectively. $g^{t\phi}$ and $g^{tt}$ are raised components of the background Kerr metric $g_{\mu \nu}$.

In a typical extreme-mass-ratio simulation, the most important code outputs are the waveform reaching the horizon (the left most point of the computational domain), the waveform reaching future null infinity (the right most point of the computational domain), and self-force quantities computed \red{near} the location of the smaller black hole $r_{*} = r_{*,p}$. We now consider our DG method's ability to compute \red{waveforms, fluxes, and the dissipative part of the self-force for given number of harmonic modes.} 

\subsubsection{Waveform and fluxes}
\label{sec:waveform_and_fluxes}

For our purposes, we study quantities computed from the waveform. In particular, we compute the energy luminosity at future null infinity ($\scri$) and the horizon $r=2M$ ($\H$) according to standard formulas~\cite{Vishal:2023fye,warburton2010self}
\begin{align} \label{eq:energy_flux}
\dot{E}^{\scri,\H}_{\ell m} = \frac{1}{4\pi} \sum_{\ell, m} \left| \partial_{t}\psi^{\scri,\H}_{\ell m} \right|^2 \,,
\end{align}
that is carried away by each multipole component $\psi^{\ell m}$. We also calculate the total energy flux by
\begin{align} \label{eq:total_energy_flux}
\dot{E}^{total}_{\ell m} =  \dot{E}^{\scri}_{\ell m} + \dot{E}^{\H}_{\ell m} \,.
\end{align}

%\begin{figure}[htbp]
\begin{figure}[t]
  \centering
  \includegraphics[width=0.5\textwidth]{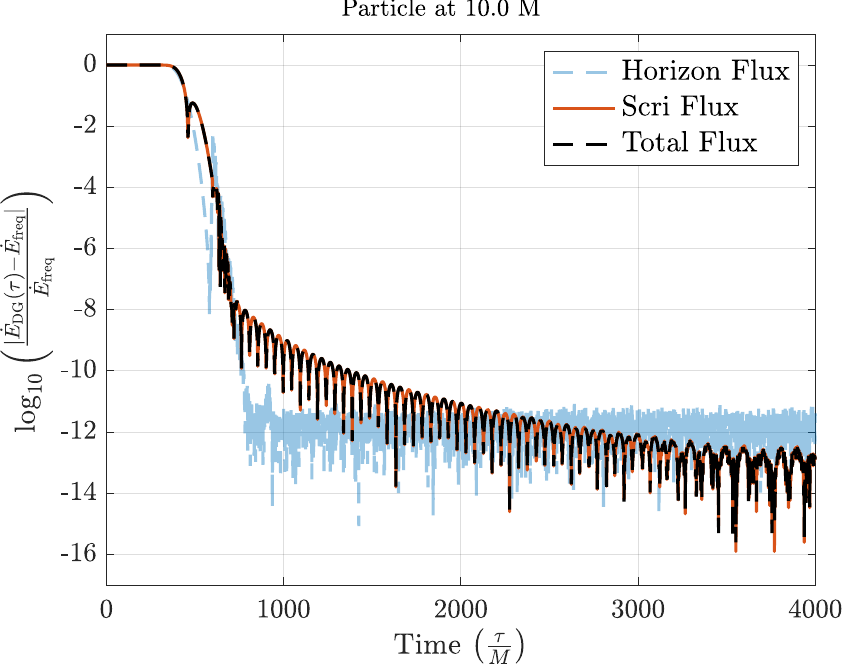}
  \caption{Energy flux error \red{(in the $(2,2)$-mode)} from our DG solver, where the fluxes are measured at both the horizon (dashed blue line) and future null infinity (solid orange line). Here the reference fluxes are computed with a frequency domain code. At early times the solution is contaminated by spurious junk due to incorrect initial data, but the solution becomes better as we wait longer. Because tails at $\mathscr{I^+}$ decay more slowly, one must wait even longer (as compared to horizon measurements) for these transients to die off.
  }   
      \label{fig:energy_flux_with_time}         
\end{figure}

Before presenting our superconvergence results, it is worth noting that 
spurious ``junk'' due to trivial initial data~\footnote{\red{While trivial data is not astrophysically realistic, it remains a standard and commonly used choice in the absence of a well-established alternative.}} needs to leave the computational domain before the system settles down into a physical steady state. Fig.~\ref{fig:energy_flux_with_time}
shows that the fluxes computed at both the horizon and $\mathscr{I^+}$ do indeed approach 
their physically expected values at late times. In this experiment, we consider 
the smaller BH to be in a circular orbit of radius $r_p=10$ about a Schwarzschild black hole of mass $M=1$. We set our numerical grid around the particle such that $\rho \in [r_{*,p}-200,r_{*,p}+200]$.
We perform a high-resolution numerical computation (setting  $N=9$, $K=100$ and $\Delta t = 0.010460591$) of the 
$(2,2)$-mode energy flux $\dot{E}^{22}_{\rm DG}(\tau)$ from 
a particle in circular orbit and compare this value to the energy flux computed with a frequency-domain solver~\cite{BHPToolkit,TeukFreqCode} 
$\dot E^\scri_{\rm freq}=3.36997747060345 \times 10^{-6}$ and $\dot E^\H_{\rm freq}=9.827090755609\times 10^{-10}$.
At early times the solution is contaminated by spurious junk due to incorrect initial data,
but the solution becomes better as we wait longer. As noted in Ref.~\cite{field2010persistent}, because tails at $\mathscr{I^+}$ decay more slowly, one must wait even longer (as compared to horizon measurements) for these transients to die off. \red{These long-lived ``'junk tails'' are not artifacts of the discontinuous Galerkin method itself but rather a generic feature of the underlying PDE when incorrect initial data is evolved in time-domain solvers that include $\mathscr{I}^+$. These junk tails are distinct from the persistent junk solutions (so-called ``Jost junk'') reported in previous work~\cite{field2010persistent}, which arise from specific numerical treatments of distributional source terms that we also use here. The persistent junk solutions (which are not intrinsic to DG schemes~\cite{field2010persistent,field2022discontinuous}) can be removed through a smoothing technique as described in Appendix~\ref{appA}. In this paper we choose soure-term smoothing parameters $\delta=0.00025$ and $\tau=400$.}

We now empirically examine the convergence of energy flux quantities at the horizon and \(\mathscr{I}^+\) as the polynomial degree \(N\) and the number of elements \(K\) vary, while keeping the timestep fixed. Our results are summarized in Fig.~\ref{fig:flux_super_convergence_with_dt_by_10}, where the top row corresponds to \(\Delta t = 0.010460591\) and the bottom row to \(\Delta t = 0.0010460591\). Although the timestep difference is relevant for self-force calculations, it does not affect the energy flux computations. 

For DG methods, the numerical error for smooth solutions is typically expected to decay at the standard rate \( \propto K^{-(N+1)} \), indicating exponential convergence. However, we observe that the energy fluxes exhibit superconvergence, with rates between \(2N\) and \(2N+1\). This is consistent with Theorem~\ref{Thm:pointwise}, which predicts superconvergence when solutions -- and consequently, the energy fluxes -- are evaluated at the roots of the Radau polynomial. Additionally, in our hyperboloidal domain setup, future null infinity lies at the rightmost boundary of the final subdomain. This boundary coincides with the Radau polynomial's outflow point, which is known to yield the optimal superconvergence rate of \( 2N+1 \). This theoretical expectation is confirmed by the numerical results in Fig.~\ref{fig:flux_super_convergence_with_dt_by_10}.

\subsubsection{Temporal component of the self-force (circular orbits)}

The total energy loss carried away by the waves must be balanced by local self-forces acting on the smaller black hole~\cite{Detweiler:2009ah,Detweiler:2005kq,Barack:2009ux,Tanaka:2005ue,Poisson:2011nh}. In general, self-force computations are divergent and typically require a regularization technique~\cite{Pound:2021qin,Warburton:2010eq,Barack:2009ux,Barack:2018yvs}. For the restricted case of scalar perturbations sourced by circular orbits (which we consider here), the energy luminosity associated with waves escaping to null infinity and down the black hole are directly related to local self-force calculations performed at the particle. \red{Furthermore, unlike the radial component of the self-force, the dissipative part (related to the $F_t$ and $F_{\phi}$ components of the self-force) can be computed} without regularization~\cite{Detweiler:2008ft,Canizares:2009ay}. Using the framework provided in Ref.~\cite{castillo2018self}, adapted to our problem, the temporal component of the self-force is 
\begin{equation}
\label{eq:ft}
  F_{t}^{\ell m} =  \frac{1}{r_p u^t} Y_{\ell m} \left(\frac{\pi}{2},\phi_p (t)\right) \partial_t \psi_{\ell m}(t,r=r_p) \,,
\end{equation}
which is related to the total energy flux as
\begin{align}
  F_{t}^{\ell m} = u^t \dot E^{total}_{\ell m} \,.
\end{align}
These relationships allow us to easily test convergence rates in a self-force computation, $F_{t}^{\ell m}$, by comparing this quantity to $\dot{E}^{total}_{\ell m}$. We note that evaluation of Eq.~\eqref{eq:ft} is challenging for most time-domain solvers due to source term $S_{\ell m}$ in Eq.~\eqref{eq:teuk1} containing a Dirac delta distribution. Previous work has shown that DG methods can exactly treat the Dirac delta distribution and provide high-accuracy in the numerically computed solution even at the location of the source~\cite{field2009discontinuous,field2022discontinuous,Vishal:2023fye}. 

\begin{figure*}[htbp]
  \centering
  \includegraphics[width=\textwidth]{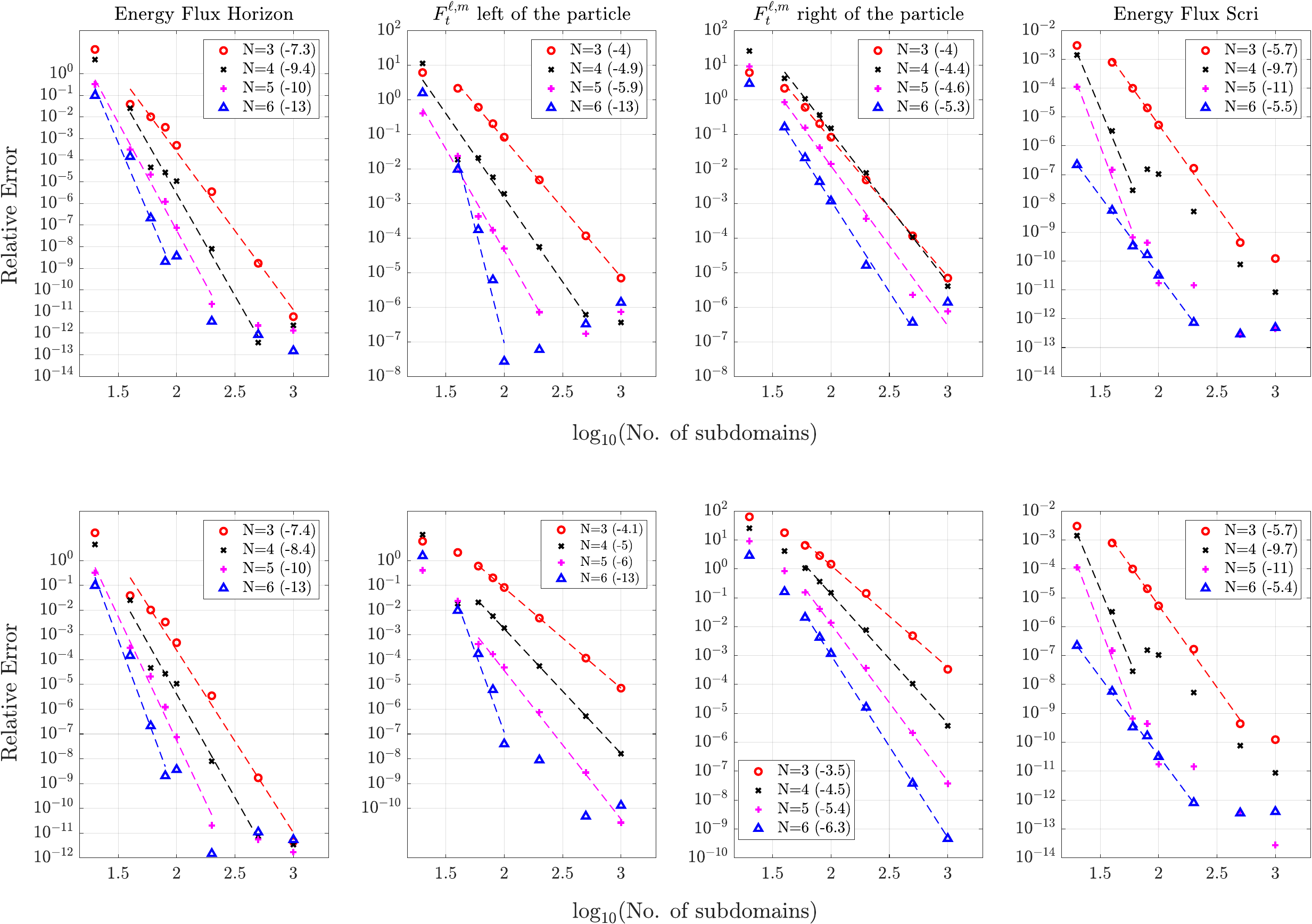}
  \caption{Convergence of energy fluxes and self-force measurements \red{$(\ell=2,m=\pm 2)$} for a particle in a circular orbit at $r_p=10$. The first and fourth columns show energy fluxes at the event horizon and future null infinity, respectively, both exhibiting superconvergence. The middle two columns display the dissipative self-force, computed at two distinct locations: one to the left of the particle and one to the right. While both self-force measurements follow the standard $\approx N+1$ convergence rate, the evaluation on the left coincides with a Radau outflow nodal point and is significantly more accurate. This behavior is examined further in Fig.~\ref{fig:outflow_inflow_with_N}. Finally, the top row corresponds to a timestep of \(\Delta t = 0.010460591\), while the bottom row uses \(\Delta t = 0.0010460591\), highlighting the greater sensitivity of self-force computations to temporal discretization errors.}
  \label{fig:flux_super_convergence_with_dt_by_10}
\end{figure*}

We now discuss our numerical experiment, where the computational and physical setup remains identical to that described in Sec.~\ref{sec:waveform_and_fluxes}. In Fig.~\ref{fig:flux_super_convergence_with_dt_by_10}, we present convergence results for the self-force computation \red{focusing on the $(\ell=2,m=\pm 2)$ component for now}. Compared to energy flux computations, achieving highly accurate self-force results requires a smaller timestep. Therefore, the bottom row of Fig.~\ref{fig:flux_super_convergence_with_dt_by_10} is more relevant when analyzing spatial convergence in the DG scheme. While we do not observe the same superconvergent rates as seen for energy fluxes (an issue currently under investigation; see appendix~\ref{appA}), we find that \( F_{t}^{\ell m} \) still converges exponentially fast. Interestingly, we notice a difference in the overall error magnitude depending on whether Eq.~\eqref{eq:ft} is evaluated to the left or right of \( r = r_p \). To further understand this difference, recall from Fig.~\ref{fig:wave_snapshot_dg} that the Dirac delta is positioned precisely at a subdomain interface. And so \( F_{t}^{\ell m} \) can be computed using the numerical solution defined on either the left or right subdomain at \( r = r_p \). When evaluating from the left subdomain, \( r = r_p \) coincides with the rightmost boundary of that subdomain, which is the Radau outflow point. While we do not observe a superconvergence rate of \( 2N+1 \), we find that extracting \( F_{t}^{\ell m} \) from the left subdomain yields orders-of-magnitude more accurate values. This effect is evident when comparing results at the same DG grid resolution (i.e., identical values of \( N \) and \( K \)) as is done in Fig.~\ref{fig:outflow_inflow_with_N}. Here the self-force error is plotted against the DG polynomial degree \( N \), using \( K = 60 \) subdomains (red line) and \( K = 100 \) subdomains (black line), with a fixed timestep of \( \Delta t = 0.0010460591 \) in both cases. 

\begin{figure}[htbp]
  \centering
  \includegraphics[width=0.5\textwidth]{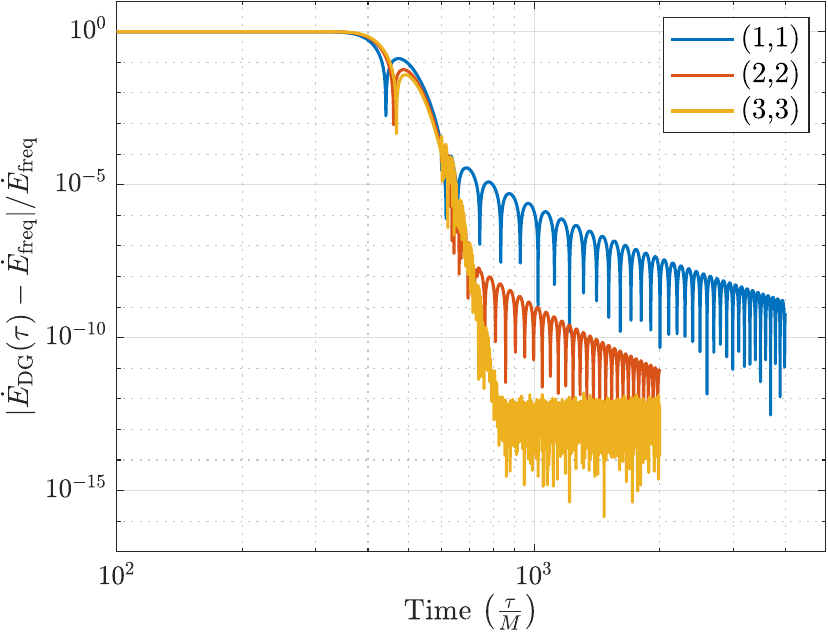}
  \caption{\red{Scri energy flux for the scalar Teukolsky equation. The energy flux is computed using Eq.~\eqref{eq:energy_flux} and is plotted as a function of time for different values of the multipole $\left(\ell,m \right)$. The energy flux is computed at the rightmost point in the grid, corresponding to null infinity using hyperboloidal layers. We observe that the energy flux converges to a constant value as time increases, indicating that the solution has settled into a steady state with machine epsilon at late time.}}
        
      \label{fig:mode_flux_error_with_time}
\end{figure}

\begin{figure}[htbp]
  \centering
  \includegraphics[width=0.5\textwidth]{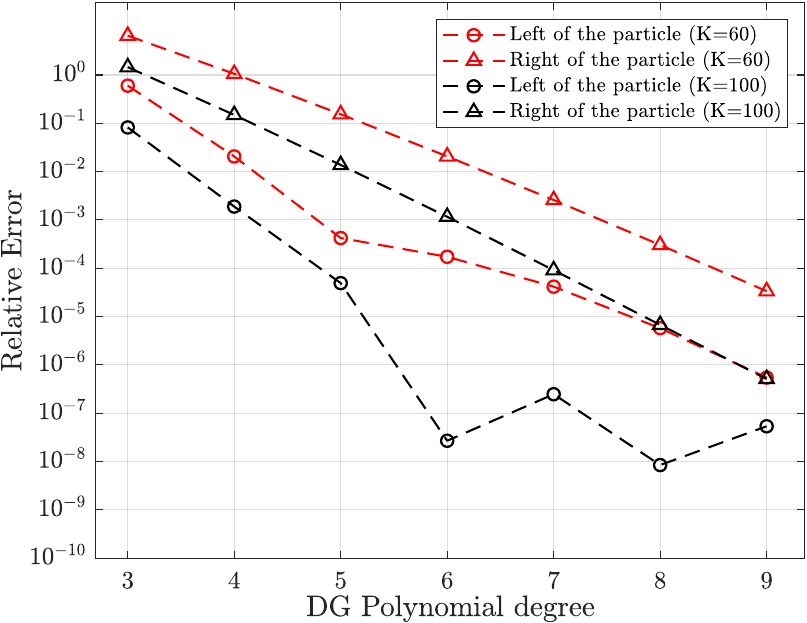}
   \caption{Relative error in the dissipative self-force $F_{t}^{22}$ for a particle in a circular orbit at $r_p = 10$. The relative error is plotted as a function of $N$ (the polynomial degree) for two grid resolutions: $K = 60$ subdomains (red line) and $K = 100$ subdomains (black line). The self-force measured to the left of the particle exhibits significantly lower error for the same  $N$ and  $K$. This behavior is likely due to the evaluation point coinciding with a Radau outflow nodal point.}
      \label{fig:outflow_inflow_with_N}
\end{figure}

\red{In Tables~\ref{tab:energy_comparison} and \ref{tab:flux_comparison} we now consider the energy fluxes and the dissipative component of the self-force using $\ell \leq 5$ modes. Using Fig.~\ref{fig:flux_super_convergence_with_dt_by_10} we find the optimal values of $N(=6)$, $K(=500)$ and $\Delta t(=0.0004)$, where $\Delta t$ was rescaled to accommodate higher m modes. Due to the fast convergence of these quantities with $\ell$, truncating the mode sum at $\ell=5$ is sufficient for obtaining accurate physical fluxes and self-force quantities (formally, the computation would require taking the limit $\ell \rightarrow \infty$). We find our DG method is able to compute accurate quantities for a range of modes. The error in certain modes could be further improved by increasing the grid resolution, but these modes are weak and their contribution to the overall mode-summed errors are small. Finally, Fig,~\ref{fig:mode_flux_error_with_time} shows the energy flux error in the $(1,1)$, $(2,2)$, and $(3,3)$ modes from our DG solver. As already discussed in Sec.~\ref{sec:waveform_and_fluxes}, junk tails contaminate the solution at early to intermediate time. As is well known, the tails decay faster for higher values of $\ell$, and so we see that the $(1,1)$ mode's junk tail requires waiting very long times to decay away. This entirely explains the larger $\approx 10^{-9}$ relative error (larger compared to the corresponding $F_t^{11}$ computation as well as energy fluxes for the other $\ell=m$ modes) in the energy flux value $E^{\infty}_{11}$ reported in Table~\ref{tab:energy_comparison}.}

\begin{table*}
  \centering
  \begin{tabular}{|c|c|c|c|c|c|c|c|}
  \hline
  $\ell$ & $m$ & $E^{\infty}_{\ell m}(BHPTK)$ & $E^{\infty}_{\ell m}(DG)$ & Rel. Error & $E^{H}_{\ell m}(BHPTK)$ & $E^{H}_{\ell m}(DG)$ & Rel. Error \\
  \hline
  \multirow{1}{*}{1} 
   & 1 & 1.1275381282792e-05  & 1.1275381227307e-05  & 4.9e-09 & 8.40476511529e-08 & 8.40476511524e-08 & 5.9e-12 \\
  \hline
  \multirow{1}{*}{2} 
   & 2 & 3.3699774706034e-06  & 3.3699774705725e-06  & 9.2e-12 & 9.82709075560e-10 & 9.82709075563e-10 & 3.1e-12 \\
  \hline
  \multirow{2}{*}{3}
   & 1 & 5.2613331734512e-11 & 5.2613331739059e-11 & 8.6e-11 & 3.03759210892e-13 & 3.03759211125e-13 & 7.7e-10 \\
   & 3 & 7.6331083748229e-07 & 7.6331083748275e-07 & 6.0e-13 & 7.25673474495e-13 & 7.25673474462e-12 & 4.5e-11\\
  \hline
  \multirow{2}{*}{4} 
   & 2 & 6.7904027182226e-11 & 6.7904027179442e-11 & 4.1e-11 & 3.67225893306e-15 & 3.67225892040e-15 & 3.4e-09 \\
   & 4 & 1.5670329789196e-07 & 1.5670329789190e-07 & 3.8e-13 & 4.58206190096e-14 &  4.58206189694e-14 & 8.8e-10 \\
  \hline
  \multirow{3}{*}{5} 
   & 1 & 3.4291108295226e-17 & 3.4291109373856e-17 & 3.1e-08 & 1.85386102570e-18 & 1.85386054661e-18 & 2.6e-07\\
   & 3 & 2.9363925683679e-11 & 2.9363925682866e-11 & 2.8e-11 & 2.76252436294e-17 & 2.76252432529e-17 & 1.4e-08 \\
   & 5 & 3.0620409217637e-08 & 3.0620409217632e-08 & 1.6e-13 & 2.69310777909e-16 & 2.69310771087e-16& 2.5e-08 \\
  \hline
  \hline
    \multicolumn{2}{|c|}{Total $(\pm m)$} & 3.1192286358612e-05 & 3.1192286247581e-05 & 3.6e-09 & 1.70075941028e-07 & 1.70075941027e-07 & 5.8e-12 \\
    \hline
  \end{tabular}
  \caption{\red{Comparison of energy values between our time-domain numerical solver (DG) and a frequency-domain solver (BHPTK). When presenting results for individual modes, we only show $m >0$ as the $m<0$ mode values are identical.}}
  \label{tab:energy_comparison}
  \end{table*}

  \begin{table*}
    \centering
    \begin{tabular}{|c|c|c|c|c|c|c|}
    \hline
    $\ell$ & $m$ & $F^{\ell,m}_{t}(BHPTK)$ & $F^{\ell,m}_{t}(DG-outflow)$ & Relative Error & $F^{\ell,m}_{t}(DG-inflow)$ & Relative Error \\
    \hline
    \multirow{1}{*}{1} 
     & 1 & 1.357711444755218e-05 & 1.357711444742770e-05 & 9.2e-12 & 1.357711442805173e-05 &  1.4e-09 \\
    \hline
    \multirow{1}{*}{2} 
     & 2 & 4.029068047679367e-06 & 4.029068047580443e-06  & 2.5e-11 & 4.029067895905402e-06 & 3.8e-08 \\
    \hline
    \multirow{2}{*}{3} 
     & 1 & 6.324802102069796e-11 & 6.324811657194986e-11 & 1.5e-06 & 6.278726094548034e-11 & 7.3e-03 \\
     & 3 & 9.123396243081638e-07 & 9.123396242444545e-07 & 7.0e-11  & 9.123389081294995e-07 &  7.8e-07 \\
    \hline
    \multirow{2}{*}{4} 
     & 2 & 8.116522516615489e-11 & 8.116493286819127e-11 & 3.6e-06 & 7.963190913052001e-11 &  1.9e-02 \\
     & 4 & 1.872963195836367e-07 & 1.872963175872844e-07 & 1.1e-08  & 1.872938444195875e-07 & 1.3e-05  \\
    \hline
    \multirow{3}{*}{5} 
     & 1 & 4.320150141576758e-17 & 9.657629096004976e-16 & 2.1e+01 & -3.609925230010641e-12 & 8.4e+04 \\
     & 3 & 3.509663707798448e-11 & 3.509926356679569e-11 & 7.5e-05 & 3.088561308671916e-11 &  1.2e-01 \\
     & 5 & 3.659838944833432e-08 & 3.659839218560491e-08 & 7.5e-08 &  3.659141866784427e-08 & 1.9e-04 \\
    \hline
    \hline
    \multicolumn{2}{|c|}{Total $(\pm m)$} & 3.748519267699629e-05 & 3.748519268460852e-05 &2.0e-10 & 3.7485152380064e-05 & 1.1e-06   \\
    \hline
    \end{tabular}
    \caption{\red{Comparison of $F_{t}^{\ell,m}$ values between our time-domain numerical solver (DG) and a frequency-domain solver (BHPTK). When presenting results for individual modes, we only show $m >0$ as the $m<0$ mode values are identical. The outflow values correspond to the values calculated to the left of the particle, while the inflow values are calculated to the right of the particle. As discussed in the main text,  data computed at outflow locations on the grid benefit  from superconvergence.}}
    \label{tab:flux_comparison}
    \end{table*}

\section{Summary}
\label{sec:summary}

In this work, we have presented a superconvergent discontinuous Galerkin (DG) method for solving the scalar Teukolsky equation on hyperboloidal domains. Unlike standard DG methods, which typically converge at a rate of $\mathcal{O}(\Delta x^{N+1})$, our approach achieves superconvergence, attaining a rate of $\mathcal{O}(\Delta x^{2N+1})$ at specific grid points corresponding to Radau nodal locations. This improved convergence rate is theoretically explained through an analysis of polynomial interpolation properties and the careful selection of numerical fluxes, which together ensure that special points on the computational grid -- such as the particle's location, the event horizon, and future null infinity -- exhibit significantly improved accuracy and convergence rates.  

By combining our DG method with hyperboloidal layer compactification, we achieve highly accurate waveforms (and derived quantities such as energy fluxes) on relatively coarse grids. We also compute the dissipative self-force for circular orbits, demonstrating further accuracy gains in these calculations. To our knowledge, this represents the first spectrally convergent self-force computation using a time-domain solver, albeit for the simplified case of circular orbits and scalar perturbations.

Future work will focus on extending these techniques to more realistic scenarios, such as eccentric or inspiraling orbits and gravitational perturbations of Kerr spacetime. While our primary motivation has been gravitational waveform modeling for extreme mass-ratio inspirals, superconvergent DG methods -- particularly when combined with hyperboloidal layers -- may have broader applications in computational relativity.

\section*{Acknowledgments}
We thank Som Bishoyi, Gaurav Khanna, John Driscoll, \red{Barry Wardell, and Anıl Zenginoğlu} for helpful discussions throughout the project.
This work was partially funded by the Swedish Research Council (VR grant 2022-03528) and was done in connection with Digital Futures and the Linné Flow Centre at KTH. Some of the simulations were performed on the UMass-URI
UNITY supercomputer supported by the Massachusetts Green High Performance Computing Center (MGHPCC). We also thank ChatGPT for help \red{doing an initial literature survey of the topics covered in the introduction.} The authors acknowledge support of NSF Grant DMS-2309609 (S.F, S.G, and M.V.), Office of Naval Research/Defense University Research Instrumentation Program (ONR/DURIP) Grant No. N00014181255 (S.F). SG's research was supported in part by NSF Grant  DMS-2309609, AFOSR Grant No. FA9550-23-1-0037, DOE Grant No. DE-SC0023164 Subaward RC114586.

\appendix

\begin{figure*}[htbp]
  \centering
  \includegraphics[width=0.8\textwidth]{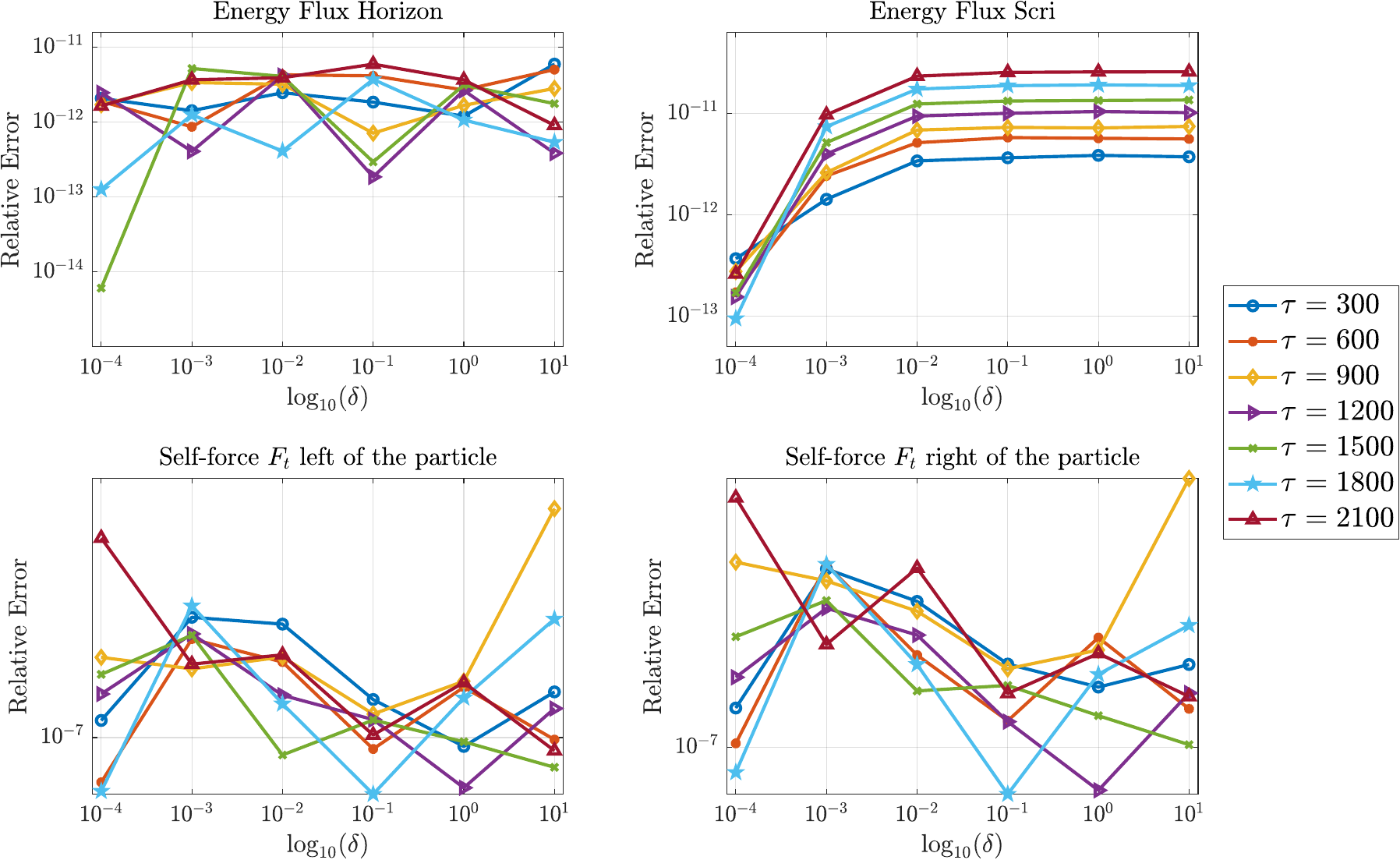}
  \caption{Effect of the source-smoothing function on energy fluxes and self-force quantities. The smoothing function, used to slowly introduce the distributional source terms, is varied by adjusting its parameters $\delta$ and $\tau$. The DG scheme parameters are fixed at $N = 9$, $K = 100$, and $\Delta t = 0.010460591$. We find that the smoothing parameters primarily affect the energy fluxes at future null infinity, while having little impact on the other quantities. This may be due to the smoothing function influencing the excitation of long-lived Price tails differently depending on the parameter choices. In particular, spurious junk tails at $\mathscr{I^+}$ decay more slowly, which could explain the observed effect.}
  \label{fig:smoothing_delta_tau}
\end{figure*}

\section{Dependence of physical quantities on the source smoothing function}
\label{appA}

In Sec.~\ref{sec:waveform_and_fluxes} we considered the distributionally sourced scalar Teukolsky equation with trivial initial data. For this problem, one often smoothly turns on the source term to mitigate transient and long-lived spurious ``junk''~\cite{field2010persistent}. The smoothing function
\begin{align*}
  S_{\ell m}(t)\to & S_{\ell m}(t)\times\\
   & \left\{ \begin{array}{ll}
  \frac{1}{2}\left[\operatorname{erf}\left(\sqrt{\delta}(t-t_{0}-\frac{\tau}{2})\right)+1\right] & \text{for }t_{0}\leq t\leq t_{0}+\tau,\\
  1 & \text{for }t>t_{0}+\tau.
  \end{array}\right .\\
  \\
\end{align*}
is parameterized by $\delta$ and $\tau$. In this appendix, we discuss the smoothing function's parameters $\delta$ and $\tau$ to observe its effect on the numerically computed energy fluxes and self-force. This study was used to find good parameters for the numerical experiments of Sec.~\ref{sec:waveform_and_fluxes}. We observe that the smoothing parameters only effect the future null infinity energy fluxes and not the other three calculations. The results are shown in Fig.~\ref{fig:smoothing_delta_tau}. 

\bibliography{References}

\end{document}